# Hall-mediated magnetic reconnection and onset of plasmoid instability.

G. Vekstein[1)] and K. Kusano[2)]

[1)]Jodrell Bank Centre for Astrophysics, University of Manchester, Manchester M13 9PL, United Kingdom

[2)]Institute for Space-Earth Environmental Research, Nagoya University, Nagoya, Aichi 464-8601, Japan

We investigate a role of the Hall-effect in the current sheet evolution and onset of the secondary tearing (plasmoid) instability in the framework of the incompressible resistive Hall-magnetohydrodynamics (MHD). The model under consideration is a force-free modification of the Taylor's problem. Thus, the first part of the paper is devoted to a detailed analytical study of the Hall-MHD forced magnetic reconnection in a tearing stable force-free magnetic configuration. Then, in the second part, these results are used to investigate when and how the plasmoid instability can develop in the course of this process.

## I.     Introduction

Magnetic reconnection, which is a change in connectivity of magnetic field lines in a highly conducting fluid, plays a crucial role in various phenomena occurring in space and laboratory plasmas (solar flares, magnetospheric substorms, tokamak disruptions, etc). The primary focus of modern magnetic reconnection research has been to explain why observed rate of reconnection is usually much faster than predicted by conventional magnetohydrodynamic (MHD) models. Therefore, until recently, the emphasis has been shifted beyond the standard MHD description. The idea was that some kinetic effects (anomalous resistivity, gyro-viscosity, Hall-effect, electron inertia, etc.) can accelerate reconnection process to such an extent that its time-scale becomes effectively independent on the Lundquist number of a system (what is now called "fast reconnection").

A renewed interest in simple MHD models has been ignited by a realization that highly elongated current sheets, which are formed in the process of magnetic reconnection under a large Lundquist number, cannot persist. They are subjected to the so-called plasmoid (secondary tearing) instability[1,2], which breaks the initially long current sheet into a chain of magnetic islands (plasmoids), whose subsequent nonlinear evolution paves the way to fast reconnection[3-5]. A large body of numerical simulations indicates that such an instability occurs when the Lundquist number exceeds some critical threshold $S_c \approx 10^4$. Since for a vast majority of applications the actual Lundquist number is by far larger than $10^4$, reaching fast reconnection via plasmoid instability looks like quite a generic scenario. Therefore, the issue of the plasmoid-mediated fast reconnection already attracted a significant number of publications. Most of them are numerical simulations, because self-consistent description of this process is not a simple task: it requires following an entire current sheet evolution which at some point brings about onset of the secondary tearing instability[6].

However, in order to get more complete understanding of the issue, in particular, its scaling with the plasma and magnetic field parameters, one needs some tractable analytical model. In this respect, it is useful to explore a well-known Taylor's model of forced magnetic reconnection, first considered in a seminal paper by Hahm and Kulsrud[7] (hereafter HK). Thus, a recent theory[8] provides detailed analytical description of nonlinear forced reconnection and onset of plasmoid instability in the framework of the standard resistive MHD. Such a scheme, however, is not applicable if the Lundquist number is so large that current sheet thickness becomes comparable to the ion inertial skin depth[9,10] $d_i = c/\omega_{pi}$. In this case flows of electrons and ions inside the current sheet are separated, which manifests itself as Hall-effect. Thus, here our goal is to investigate when and how Hall-effect changes the pace of forced magnetic reconnection and appearance of the plasmoid instability.

In what follows, we use the force-free modification[11] of the Taylor's model, when uniform plasma with the initial magnetic field

$$\vec{B}^{(0)} = (0, B_0 \sin \alpha x, B_0 \cos \alpha x), \quad (1)$$

is confined between the two perfectly conducting boundaries located at $x = x_b = \pm a$. This equilibrium is subjected to a boundary deformation as

$$x_b = \mp(a + \delta \cos ky), \quad (2)$$

with $\frac{\delta}{a} \ll 1$. In the linear approximation with respect to this small parameter, there are two new equilibria consistent with the deformed boundaries, which in terms of the flux-function $\Psi(x, y,)$ can be written as

$$\vec{B}^{(i,r)} = (\vec{\nabla}\Psi^{(i,r)} \times \hat{z}) + \alpha\Psi^{(i,r)}\hat{z}, \quad \Psi^{(i,r)} = \Psi^{(0)}(x) + \psi_1^{(i,r)}(x)\cos ky. \qquad (3)$$

Here $\Psi^{(0)}(x) = \frac{B_0}{\alpha}\cos\alpha x$ corresponds to the initial magnetic field (1), while two functions related to the perturbation are equal to

$$\psi^{(i)}(x) = B_0\delta\frac{\sin\alpha a}{\sin\kappa a}|\sin\kappa x|, \quad \psi^{(r)}(x) = B_0\delta\frac{\sin\alpha a}{\cos\kappa a}\cos\kappa x. \qquad (4)$$

It is assumed that $\kappa^2 = \alpha^2 - k^2 > 0$ (as explained below, long-wave perturbations with $k \ll \alpha$ are most important). The first above-given solution, $\psi^{(i)}(x)$, represents the ideal MHD perturbed equilibrium, which preserves topology of the initial field (1) but acquires discontinuity at $x = 0$: the magnetic field component $B_y = -\partial\Psi/\partial x$ has there a finite jump:

$$\{B_y\} \equiv B_y\big|_{0+} - B_y\big|_{0-} = -2B_0\kappa\delta\frac{\sin\alpha a}{\sin\kappa a}. \qquad (5)$$

On the other hand, the solution $\psi^{(r)}$ is a regular one, but topology of the respective equilibrium differs from that of the initial filed: magnetic field lines reconnect and form magnetic islands located at the plane $x = 0$. The magnetic flux confined inside a single island is equal to

$$\Delta\psi^{(r)} = 2\psi^{(r)}(x=0) = 2B_0\delta\frac{\sin\alpha a}{\cos\kappa a} \qquad (6)$$

As demonstrated by HK, forced reconnection is a process of the transition from the ideal MHD equilibrium to the reconnected one, which takes place when a small but finite plasma resistivity is present.

Forced reconnection results in a release of the excess (free) magnetic energy stored in the initial magnetic configuration. Indeed, following Ref.11 , the magnetic energy [per unit area in the ($x - y$) plane ] for the ideal MHD equilibrium is equal to $W_M^{(i)} = W_M^{(0)} + \Delta W_M^{(i)}$, where $W_M^{(0)} = 2aB_0^2/8\pi$ is the energy of the initial field, and

$$\Delta W_M^{(i)} = \frac{B_0^2 \delta^2 \sin^2(\alpha a)}{8\pi a}[(\kappa a)\cot(\kappa a) - (\alpha a)\cot(\alpha a)] > 0 \qquad (7)$$

Thus, such external perturbation moves the magnetic energy slightly up. On the other hand, for the terminal reconnected state one gets $W_M^{(r)} = W_M^{(0)} + \Delta W_M^{(r)}$, with

$$\Delta W_M^{(r)} = -\frac{B_0^2 \delta^2 \sin^2(\alpha a)}{8\pi a}[(\kappa a)\tan(\kappa a) + (\alpha a)\cot(\alpha a)] < 0 \qquad (8)$$

Therefore, the total magnetic energy released in the process of forced reconnection reads

$$\Delta W_M = W_M^{(i)} - W_M^{(r)} = \frac{B_0^2 \delta^2 \sin^2(\alpha a)}{8\pi a}(\kappa a)[\cot(\kappa a) + \tan(\kappa a)] > 0 \qquad (9)$$

As seen from (7-9), the released energy exceeds the extra magnetic energy supplied by external perturbation, since the final reconnected state has lower magnetic energy than the initial field. Thus, forced reconnection can be viewed as a mechanism of internal magnetic relaxation[11]. Moreover, a difference between the supplied and released energies can be very large, if the initial magnetic configuration is only marginally stable, i.e. its parameters are close to the MHD instability threshold. Clearly, the very issue of forced reconnection makes sense only when the initial state is MHD stable. It is well-known that the field (1) explored here is stable in the framework of the ideal MHD, and becomes tearing unstable if $\kappa a > \pi/2$ (see, e.g., Ref.11]. Thus, modes with $k \to 0$ are most unstable, which yields the instability threshold $\varepsilon \equiv \alpha a > \varepsilon_{cr} = \pi/2$. When parameter $\kappa a$ approaches the instability threshold, the released magnetic energy given by Eq.(9) becomes formally divergent. The reason is that at this limit the size of the magnetic island [see Eq.(6) for its magnetic flux] is also growing to infinity. This indicates that the linear approximation used so far is not applicable. Taking into account a finite size of magnetic islands saturates the released energy[12], which, however, still greatly exceeds the extra energy (7) provided by the external source[13]. Therefore, the latter is energetically insignificant, and its role is just to trigger the internal relaxation process. In what follows our interest is with the dynamics of forced reconnection, which is determined by evolution of the central current sheet at $x = 0$ and, hence, is not sensitive to the nonlinearity of the external solution. Thus, we assume that the system is not close to the instability threshold, the respective parameter $\varepsilon \sim 1$, and expressions (4) for the perturbed equilibria hold.

If plasma thermal pressure is not too small (plasma $\beta \equiv 8\pi P / B_0^2 > S^{-2/5}$, where $S \equiv aV_A / \eta \gg 1$ is the relevant Lundquist number), in the Hall-MHD magnetic reconnection the plasma flow may be considered as incompressible[14]. Then, by representing the magnetic field and the plasma velocity as

$$\vec{B}(x,y,t) = \vec{\nabla}\Psi(x,y,t) \times \hat{z} + B_z(x,y,t)\hat{z}, \quad \vec{V}(x,y,t) = \vec{\nabla}\Phi(x,y,t) \times \hat{z} + V_z(x,y,t)\hat{z},$$

where $\Phi$ is a stream-function of the flow in the $(x-y)$ plane, equations of motion for $\Phi$ and $V_z$ take the form

$$\rho \frac{d}{dt}(\nabla^2 \Phi) = \frac{1}{4\pi}[\vec{\nabla}\Psi \times \vec{\nabla}(\nabla^2 \Psi)] \cdot \hat{z}, \tag{10}$$

$$\rho \frac{dV_z}{dt} = \frac{1}{4\pi}(\vec{\nabla}B_z \times \vec{\nabla}\Psi) \cdot \hat{z} \tag{11}$$

These should be complemented with the Maxwell's equations

$$\frac{\partial \vec{B}}{\partial t} = -c(\vec{\nabla} \times \vec{E}), \quad \vec{j} = \frac{c}{4\pi}(\vec{\nabla} \times \vec{B}), \tag{12}$$

where the electric field $\vec{E}$ is equal to

$$\vec{E} = -\frac{1}{c}(\vec{V}_e \times \vec{B}) + \frac{1}{\sigma}\vec{j} = -\frac{1}{c}(\vec{V} \times \vec{B}) - \frac{1}{4\pi ne}[\vec{B} \times (\vec{\nabla} \times \vec{B})] + \frac{c}{4\pi\sigma}(\vec{\nabla} \times \vec{B}) \tag{13}$$

It follows then from (12) and (13) that

$$\frac{\partial \Psi}{\partial t} = (\vec{\nabla}\Psi \times \vec{\nabla}\Phi) \cdot \hat{z} + \eta \nabla^2 \Psi + \frac{c}{4\pi ne}(\vec{\nabla}\Psi \times \vec{\nabla}B_z) \cdot \hat{z}, \tag{14}$$

$$\frac{\partial B_z}{\partial t} = (\vec{\nabla}B_z \times \vec{\nabla}\Phi) \cdot \hat{z} - (\vec{\nabla}V_z \times \vec{\nabla}\Psi) \cdot \hat{z} + \eta \nabla^2 B_z + \frac{c}{4\pi ne}(\vec{\nabla}\nabla^2\Psi \times \vec{\nabla}\Psi) \cdot \hat{z}, \tag{15}$$

with $\eta \equiv c^2 / 4\pi\sigma$ being plasma magnetic viscosity.

Thus, our analysis of the Hall-MHD forced magnetic reconnection is based on Eqs.(10-11) and (14-15), and the paper is organized as follows. Section II is devoted to the linear theory, results of which are then used in Section III for demonstrating how onset of the plasmoid instability is affected by inclusion of the Hall-effect. A brief summary of the results and discussion are presented in Section IV.

## II. Linear regime of the Hall-MHD forced reconnection

In the linear approximation the governing Eqs. (10-15) take the form:

$$\frac{\partial}{\partial t}\left(\nabla^2 \Phi\right) = -\frac{1}{4\pi\rho}\left[\frac{d\Psi^{(0)}}{dx}\cdot\frac{\partial}{\partial y}\left(\nabla^2 \Psi_1\right) - \frac{d^3\Psi^{(0)}}{d^3 x}\cdot\frac{\partial \Psi_1}{\partial y}\right], \qquad (16a)$$

$$\frac{\partial V_z}{\partial t} = \frac{1}{4\pi\rho}\left(\frac{dB_z^{(0)}}{dx}\cdot\frac{\partial \Psi_1}{\partial y} - \frac{d\Psi^{(0)}}{dx}\cdot\frac{\partial B_z^{(1)}}{\partial y}\right), \qquad (16b)$$

$$\frac{\partial \Psi_1}{\partial t} = -\frac{d\Psi^{(0)}}{dx}\cdot\frac{\partial \Phi}{\partial y} + \eta\nabla^2 \Psi_1 + \frac{c}{4\pi n e}\left(\frac{d\Psi^{(0)}}{dx}\cdot\frac{\partial B_z^{(1)}}{\partial y} - \frac{dB_z^{(0)}}{dx}\cdot\frac{\partial \Psi_1}{\partial y}\right), \quad (16c)$$

$$\frac{\partial B_z^{(1)}}{\partial t} = -\frac{dB_z^{(0)}}{dx}\cdot\frac{\partial \Phi}{\partial y} - \frac{d\Psi^{(0)}}{dx}\cdot\frac{\partial V_z}{\partial y} + \eta\nabla^2 B_z^{(1)} + \frac{c}{4\pi n e}\left(\frac{d^3\Psi^{(0)}}{d^3 x}\cdot\frac{\partial \Psi_1}{\partial y} - \frac{d\Psi^{(0)}}{dx}\cdot\frac{\partial}{\partial y}\nabla^2 \Psi_1\right),$$
(16d)

where $\Phi, V_z, \Psi_1, B_z^{(1)}$ are perturbations proportional to the first power of a small parameter $(\delta/a)$. Since reconnection takes place inside a narrow central current sheet with a thickness $\Delta \ll a$ (as well as $\Delta \ll k^{-1}$ - a wavelength of the external boundary perturbation), in what follows one can simplify $\frac{dB_z^{(0)}}{dx}, \frac{d\Psi^{(0)}}{dx} \propto \sin\alpha x$ as $\propto \alpha x$, and assume that $\nabla^2(\Psi_1, \Phi, B_z^{(1)}) \approx \partial^2(\Psi_1, \Phi, B_z^{(1)})/\partial^2 x$. Furthermore, it is useful to introduce non-dimensional variables by scaling all lengths with $a$, time with the Alfven time-scale $\tau_A = a/V_A = a\sqrt{4\pi\rho}/B_0$, and for perturbations: velocity $V_z$ – with the Alfven speed as $V_A(\delta/a)$, stream-function $\Phi$ - with $aV_A(\delta/a) = V_A\delta$, flux-function $\Psi_1$ - with $aB_0(\delta/a) = B_0\delta$, and $B_z^{(1)}$ - with $B_0(\delta/a)$. These transform Eqs.(14) into the following:

$$\frac{\partial}{\partial t}\left(\frac{\partial^2 \Phi}{\partial^2 x}\right) \approx \varepsilon x \frac{\partial}{\partial y}\left(\frac{\partial^2 \Psi_1}{\partial^2 x}\right), \tag{17a}$$

$$\frac{\partial V_z}{\partial t} \approx \varepsilon x \frac{\partial}{\partial y}\left(B_z^{(1)} - \varepsilon\Psi_1\right), \tag{17b}$$

$$\frac{\partial \Psi_1}{\partial t} \approx \varepsilon x \frac{\partial \Phi}{\partial y} + \frac{1}{S}\frac{\partial^2 \Psi_1}{\partial^2 x} - \varepsilon d_i x \frac{\partial}{\partial y}\left(B_z^{(1)} - \varepsilon\Psi_1\right), \tag{17c}$$

$$\frac{\partial B_z^{(1)}}{\partial t} \approx \varepsilon^2 x \frac{\partial \Phi}{\partial y} + \frac{1}{S}\frac{\partial^2 B_z^{(1)}}{\partial^2 x} + \varepsilon \frac{\partial V_z}{\partial y} + \varepsilon d_i x \frac{\partial}{\partial y}\left(\frac{\partial^2 \Psi_1}{\partial^2 x}\right), \tag{17d}$$

where $S \equiv \frac{aV_A}{\eta}$ is the Lundquist number, and $d_i \equiv \frac{c}{a\omega_{pi}}$ is the scaled ion inertial length. In what follows it is assumed that $S \gg 1$ and $d_i \ll 1$, which is the case for a vast majority of applications.

Consider now symmetry properties of perturbations. As far as the flux-function $\Psi_1$ is concerned, that is imposed by the boundary deformation (2): $\Psi_1(x,y,t) = \psi(x,t)\cos ky$, with $\psi(x,t)$ being an even function of $x$. Then, according to (17a), $\Phi(x,y,t) = \phi(x,t)\sin ky$, where $\phi$ is an odd function of $x$. The magnetic field component $B_z^{(1)}$ is, according to (17d), a superposition of both modes:

$$B_z^{(1)}(x,y,t) = b_1(x,t)\cos ky + b_2(x,t)\sin ky, \tag{18}$$

where $b_1$ and $b_2$ are, respectively, even and odd functions of $x$. Appearance of the latter is entirely due to the Hall-effect: the second term on the r.h.s. of Eq.(18) represents a quadrupole magnetic structure which is a signature of the Hall-mediated magnetic reconnection[15]. Then, a straightforward inspection of Eqs. (17c) and (17d) reveals that $b_1(x,t) = \varepsilon\psi(x,t)$, so, according to (17c), this part of $B_z^{(1)}$ does not affect evolution of the flux-function $\Psi_1$. Finally, Eq.(17b) yields $V_z(x,y,t) = v(x,t)\cos ky$, where $v$ is an even function of $x$, and one gets the following set of evolution equations for the above-introduced functions $\psi, \phi, b_2, v$:

$$\frac{\partial}{\partial t}\phi'' = -\varepsilon k x \psi'', \tag{19a}$$

$$\frac{\partial v}{\partial t} = \varepsilon k x b_2, \tag{19b}$$

$$\frac{\partial \psi}{\partial t} = \varepsilon k x \phi + \frac{1}{S}\psi'' - \varepsilon k d_i x b_2, \tag{19c}$$

$$\frac{\partial b_2}{\partial t} = -\varepsilon k x v + \frac{1}{S}b_2'' - \varepsilon k d_i x \psi''. \tag{19d}$$

The last terms on the r.h.s. of Eqs. (19c) and (19d), which are proportional to the parameter $d_i$, are due to Hall-effect. The limit $d_i = 0$ corresponds to the standard single-fluid MHD when $b_2 = v = 0$. Thus, we aim to derive a threshold value of $d_i$, above which Hall effect makes a difference, and to investigate the resulting process of the Hall-mediated forced magnetic reconnection.

As demonstrated by HK, the boundary deformation (2) leads to appearance of the current sheet (CS) located around the plane $x = 0$, the thickness of which is decreasing with time. Under the condition $d_i \ll 1, S \gg 1$, an initial stage of this process can be described in terms of the ideal single-fluid MHD as follows. Let $\Delta(t)$ be the thickness of this CS, so that at $x \leq \Delta$ one gets

$$B_y^{(1)} \sim \frac{x}{\Delta} \Rightarrow \psi_i \sim \frac{x^2}{\Delta} \sim \Delta \Rightarrow \frac{\partial \psi_i}{\partial t} \sim \frac{d\Delta}{dt}, \psi_i'' \sim \frac{1}{\Delta}, \tag{20}$$

(a symbol $\psi_i$ here indicates the ideal MHD flux-function). It follows then from (19a) that

$$\frac{\partial}{\partial t}\phi'' \sim \frac{\partial}{\partial t}\left(\frac{\phi}{\Delta^2}\right) \sim \varepsilon k \Rightarrow \frac{\phi}{\Delta^2} \sim \varepsilon k t. \tag{21}$$

On the other hand, Eq.(19c) yields

$$\frac{d\Delta}{dt} \sim \varepsilon k \Delta \cdot \varepsilon k t \Delta^2 \Rightarrow \Delta(t) \sim (\varepsilon k t)^{-1}, \tag{22}$$

in accordance with HK. Thus, such shrinking of the CS would bring about (though only asymptotically in time) the singular ideal MHD equilibrium given by Eqs. (3) and (4). However, this process comes to the end when, eventually, a finite plasma resistivity or the Hall-effect intervene. Consider first a role of the resistivity. The respective term in Eq.(19c) can be estimated as $S^{-1}\psi_i'' \sim S^{-1}\Delta^{-1} \sim S^{-1}\varepsilon kt$, and it becomes comparable with

$$\frac{\partial \psi_i}{\partial t} \sim \frac{d\Delta}{dt} \sim (\varepsilon k)^{-1} t^{-2} \text{ at } t \sim t_S \sim (\varepsilon k)^{-2/3} S^{1/3}. \qquad (23)$$

A similar derivation for the Hall-effect is as follows. One can use Eqs. (19d) and (20) to estimate generation in the CS of the quadrupole field $b_2$:

$$\frac{\partial b_2}{\partial t} \sim \varepsilon k d_i \Rightarrow b_2 \sim \varepsilon k d_i t \text{ , so the Hall term in Eq.(19c) is of the order of}$$

$\varepsilon k d_i x b_2 \sim \varepsilon k d_i^2$. By comparing it with the l.h.s. term $\frac{\partial \psi_i}{\partial t} \sim \frac{d\Delta}{dt} \sim (\varepsilon k)^{-1} t^{-2}$, one concludes that the Hall-effect comes into play at

$$t \sim t_H \sim (\varepsilon k d_i)^{-1}, \qquad (24)$$

when, according to (22), the CS thickness $\Delta(t \sim t_H) \sim d_i$.

Thus, an interplay between the two effects depends on the relation between the $t_S$ and $t_H$. Consider first the case when the resistivity comes first, i.e. $t_S \ll t_H$, which implies that

$$d_i \ll (\varepsilon k)^{-1/3} S^{-1/3}. \qquad (25)$$

It turns out that in this case the Hall-effect does not play any role at all in forced reconnection.

In order to demonstrate this, consider what happens at $t > t_S$, when, according to HK, the system evolves in the so-called "constant-$\Psi$" regime[16]. Indeed, the amount of reconnected magnetic flux, $\psi_r$, is equal to $\psi(x=0)$, hence, as it follows from Eqs. (19c), (20) and (22), at $t \leq t_S$

$$\frac{d\psi_r}{dt} = \frac{1}{S}\psi_i'' \sim S^{-1}\Delta^{-1} \sim S^{-1}\varepsilon kt \Rightarrow \psi_r \sim S^{-1}\varepsilon kt^2 .$$

Thus, at $t \sim (\varepsilon k)^{-2/3} S^{1/3} \sim t_S$ the reconnected flux becomes comparable to the total variation of the magnetic flux function inside the CS, which is given by $\Delta \psi = \psi_i(x \sim \Delta) - \psi_i(0) = \psi_i(\Delta) \sim \Delta \sim (\varepsilon k t)^{-1}$. Therefore, at $t > t_S$ the "constant-$\Psi$" approximation holds, and temporal evolution of the reconnected flux $\psi_r(t)$ and the CS thickness $\Delta(t)$ can be obtained in the following way. First, as long as reconnected flux is still small compared to its terminal value, i.e. $\psi_r \ll 1$, the discontinuity of $B_y^{(1)}$ across the CS persists, so $\psi'' \cdot \Delta \sim 1$. Second, in this regime convective and resistive terms in Eq.(19c) should be comparable, which yields
$S^{-1}\psi'' \sim S^{-1}\Delta^{-1} \sim \varepsilon k x \phi \sim \varepsilon k \Delta \phi \Rightarrow \phi \sim S^{-1}(\varepsilon k)^{-1}\Delta^{-2}$. By inserting this expression for $\phi$ into Eq.(19a), one gets:

$$\frac{\partial}{\partial t}\phi'' \sim \frac{\partial}{\partial t}\left(\frac{\phi}{\Delta^2}\right) \sim \varepsilon k x \psi'' \sim \varepsilon k \Delta \psi'' \sim \varepsilon k \Rightarrow \Delta \sim S^{-1/4}(\varepsilon k)^{-1/2} t^{-1/4}. \tag{26}$$

As seen from (26), the CS shrinking continues, and, according to (19c), it yields the reconnection rate

$$\frac{d\psi_r}{dt} \sim \frac{1}{S}\psi'' \sim \frac{1}{S\Delta} \sim S^{-3/4}(\varepsilon k)^{1/2} t^{1/4} \Rightarrow \psi_r \sim S^{-3/4}(\varepsilon k)^{1/2} t^{5/4}. \tag{27}$$

Thus, it follows from (27) that the reconnected flux becomes of the order of unity at

$$t \approx \tau_r \sim S^{3/5}(\varepsilon k)^{-2/5}, \tag{28}$$

which, in accordance with HK, is the standard MHD reconnection time.

Now one can estimate the magnitude and, hence, significance of the ignored so far Hall term in Eq.(19c). In order to do so, it is necessary first to evaluate the quadrupole field $b_2$ generated in the CS by the Hall-effect [see Eq.(19d)]. It turns out that at $t > t_S$ the respective last term on the r.h.s. of (19b) is balanced by the resistive diffusion of $b_2$, hence

$$\varepsilon k d_i x \psi'' \sim \varepsilon k d_i \sim \frac{b_2''}{S} \sim \frac{b_2}{S\Delta^2} \Rightarrow b_2 \sim \varepsilon k d_i S \Delta^2 \sim d_i (S/t)^{1/2}.$$

Inserting this expression for $b_2$ into the Hall term in Eq.(19c), one gets

$\varepsilon k d_i x b_2 \sim \varepsilon k d_i \Delta b_2 \sim (\varepsilon k)^{1/2} d_i^2 S^{1/4}/t^{3/4}$, which at $t > t_S$ is small compared to other terms in this equation. Indeed, its ratio to $d\psi_r/dt$ [see Eq.(27)] reads

$$\frac{d_i^2 S}{t} \sim \left(\frac{d_i}{(\varepsilon k)^{-1/3} S^{-1/3}}\right)^2 \cdot \frac{t_S}{t} \ll 1 \text{ under condition (25)}.$$

In this context it is worth noting that, contrary to common wisdom, the Hall- effect could remain insignificant even when in a course of reconnection the CS thickness $\Delta$ gets smaller than $d_i$. Indeed, according to (26) and (28), $\Delta(t \sim \tau_r) \sim (\varepsilon k)^{-2/5} S^{-2/5}$, which could be smaller than $d_i$ even under the constraint (25).

Thus, the Hall-mediated regime of forced reconnection requires (at least, as shown below) the inequality opposite to (25):

$$d_i > d_i^{(1)} \sim (\varepsilon k)^{-1/3} S^{-1/3}, \qquad (29)$$

so that $t_H < t_S$, and the Hall-effect comes into play before a finite plasma resistivity intervenes. Therefore, in this case what initially follows at $t > t_H \sim (\varepsilon k d_i)^{-1}$ is a phase of the ideal Hall-MHD, when evolution of $\psi$ and $b_2$ is governed entirely by the Hall terms in Eqs. (19c) and (19d). It results in a further shrinking of the CS, which can be derived in the same way as explored in Eqs.(20-22). Thus, Eq. (19d) now yields

$$\frac{\partial b_2}{\partial t} \sim \varepsilon k d_i \Delta \psi'' \sim \frac{1}{t_H} \Rightarrow b_2 \sim t/t_H, \qquad (30)$$

and, by inserting it into Eq.(19c), one gets

$$\frac{\partial \psi}{\partial t} \sim \frac{d\Delta}{dt} \sim \varepsilon k d_i x b_2 \sim \Delta \frac{t}{t_H^2} \Rightarrow \Delta(t) \sim d_i \exp(-t^2/t_H^2). \qquad (31)$$

Such exponential collapse of the CS [which is much faster than that in the standard MHD, see Eq.(22)] originates from the dispersive character of the Hall-MHD waves (whistlers). This ideal phase of evolution holds until the resistivity intervenes at some time $t \sim t_*$, when the CS thickness becomes sufficiently small: $\Delta(t_*) \ll d_i$. This instant can be obtained by equating the resistive and Hall terms in Eq.(19c) as follows:

$$\frac{1}{S} \psi'' \sim \frac{1}{S\Delta(t_*)} \sim \varepsilon k d_i \Delta(t_*) b_2(t_*). \qquad (32)$$

Then, since temporal variation of $\Delta$ is, according to (31), much stronger than that of $b_2$ in (30), with a logarithmic accuracy the sought after time $t_* \sim t_H$. Therefore, $b_2(t_*) \sim 1$, and it follows than from (32) that

$$\Delta(t_*) \equiv \Delta_H \sim S^{-1/2}(\varepsilon k d_i)^{-1/2} \sim S^{-1/2} t_H^{1/2}, \tag{33}$$

[note that the anticipated inequality $\Delta_H \ll d_i$ is satisfied because of the condition (29)].

The subsequent resistive Hall-MHD reconnection is quite similar to the standard MHD case briefly discussed above, albeit advection of the magnetic field into the CS is now provided by the Hall-effect rather than by the plasma inflow. First, one can verify that reconnection proceeds now in the "constant-$\Psi$" regime. Indeed, in the course of the CS shrinking its internal magnetic flux is decreasing with time [see Eq.(20)] as $\Delta \psi = \psi_i(x \sim \Delta) - \psi_i(0) = \psi_i(x \sim \Delta) \sim \Delta$, hence $\Delta \psi(t_*) \sim \Delta_H$. On the other hand, the reconnected flux $\psi_r$ is growing with time as

$$\frac{d\psi_r}{dt} = \frac{1}{S}\psi'' \sim \frac{1}{S\Delta_H} \Rightarrow \psi_r(t_*) \sim \frac{t_*}{S\Delta_H} \sim \frac{t_H}{S\Delta_H},$$

hence, it becomes comparable to $\Delta\psi(t_*)$ under $\Delta_H$ given by Eq.(33). Thus, the set of relations governing the subsequent temporal evolution of $\Delta$, $b_2$ and $\psi_r$ is as follows:

$\Delta \cdot \psi'' \sim 1$ - discontinuity of $B_y^{(1)}$ across the CS;

$\frac{1}{S}\psi'' \sim \varepsilon k d_i x b_2 \Rightarrow \frac{1}{S\Delta} \sim \varepsilon k d_i \Delta b_2$ - balance of the resistive and Hall terms in Eq.(19c);

$\frac{1}{S}b_2'' \sim \varepsilon k d_i x \psi'' \Rightarrow \frac{b_2}{S\Delta^2} \sim \varepsilon k d_i \Delta \psi'' \sim \varepsilon k d_i$ - balance of the resistive and Hall terms in Eq.(19d).

These yield $\Delta \sim \Delta_H, b_2 \sim 1$ and

$$\frac{d\psi_r}{dt} = \frac{1}{S}\psi'' \sim \frac{1}{S\Delta} \sim S^{-1/2}(\varepsilon k d_i)^{1/2} \Rightarrow \psi_r(t) \sim S^{-1/2}(\varepsilon k d_i)^{1/2} t. \tag{34}$$

Therefore, if this regime proceeded until full completion of the process of forced reconnection, when $\psi_r \approx 1$, the respective reconnection time, according to (34), would be equal to

$$\tau_r^{(H)} \sim S^{1/2}(\varepsilon k d_i)^{-1/2}. \tag{35}$$

Note that the scaling (35) does not involve the ion mass $m_i$ (both $S$ and $d_i$ are proportional to $\sqrt{m_i}$), so it corresponds to the electron-MHD limit in the theory of forced magnetic reconnection[17]

It turns out, however, that this is the case only when the Hall parameter $d_i$ exceeds a certain second threshold, $d_i^{(2)}$ (see below), which is much higher than $d_i^{(1)}$ given in Eq.(29). Otherwise, at some time, $\tilde{t} \ll \tau_r^{(H)}$, the Hall regime (34) gives way to the standard MHD reconnection, and the overall reconnection time becomes equal to $\tau_r$ defined in Eq.(28). The reason lies in a double-layer structure of the CS during the resistive phase of the Hall-MHD reconnection[18,19]. Thus, the resistive region, $x \leq \Delta_H$, is surrounded by a much wider layer, $\Delta_H < x < x_H$, where the plasma resistivity plays no role, but the poloidal magnetic field described by the flux function $\psi$ is still advected towards the reconnection site by the Hall-effect [the last term on the r.h.s. of Eq.(19c)]. Therefore, by using Eqs.(19c) and (34), one can evaluate there the required quadrupole field component as

$$b_2 = -\frac{1}{\varepsilon k d_i x}\frac{\partial \psi}{\partial t} \sim \frac{S^{-1/2}(\varepsilon k d_i)^{-1/2}}{x} \sim \frac{\Delta_H}{x}. \tag{36}$$

This field also generates, according to Eq.(19b), the $z$-component of the plasma velocity:

$$\frac{\partial v}{\partial t} = \varepsilon k x b_2 \sim \varepsilon k \Delta_H \Rightarrow v \sim \varepsilon k \Delta_H t.$$

Furthermore, this velocity is responsible for balancing the Hall term in Eq.(19d), so the electric current in this layer, $\psi''$, can be estimated as

$$\varepsilon k d_i x \psi'' \sim \varepsilon k x v \Rightarrow \psi'' \sim (\varepsilon k) d_i^{-1} \Delta_H t \sim d_i^{-2} \Delta_H (t/t_H). \tag{37}$$

This current accelerate poloidal plasma flow at the rate given by Eq.(19a):

$$\frac{\partial}{\partial t}(\phi'') \sim \varepsilon k x d_i^{-2} \Delta_H \frac{t}{t_H} \Rightarrow \phi'' \sim d_i^{-3} \Delta_H x^3 \left(\frac{t}{t_H}\right)^2 \Rightarrow \phi \sim \Delta_H \left(\frac{x}{d_i}\right)^3 \left(\frac{t}{t_H}\right)^2.$$

Finally, by inserting this expression into Eq.(19c), one can estimate the width $x_H$ of the ideal Hall sublayer: at $x \sim x_H$ advection of the magnetic field by the plasma flow becomes comparable to that by the Hall term. Hence,

$$\varepsilon k x_H \phi(x_H) \sim \varepsilon k d_i x_H b_2(x_H) \sim \varepsilon k d_i \Rightarrow x_H \sim d_i \left(\frac{t}{t_H}\right)^{-1/2}, \qquad (38)$$

(at $x > x_H$ the Hall-effect is not important, and the standard MHD description applies).

Therefore, the Hall-MHD regime of reconnection described by Eq.(34) holds as long as the width of the resistive sublayer, $\Delta_H$, is smaller than $x_H$, i.e., according to (33) and (38), $t < \tilde{t} \sim Sd_i^2$. This leaves one with two possibilities. If

$$d_i > d_i^{(2)} \sim S^{-1/5}(\varepsilon k)^{-1/5}, \qquad (39)$$

[note, that $d_i^{(2)} >> d_i^{(1)} \sim S^{-1/3}(\varepsilon k)^{-1/3}$], $\tilde{t} >> \tau_r^{(H)} \sim S^{1/2} t_H^{1/2}$, and the Hall-MHD regime has enough time to complete the reconnection process. Alternatively, when $d_i^{(1)} < d_i < d_i^{(2)}$, a transition from the Hall-MHD regime (34) to the standard MHD reconnection (26) occurs at $t \approx \tilde{t}$. At this point the amount of reconnected magnetic flux is still small, $\psi_r(\tilde{t}) \sim (d_i/d_i^{(2)})^{5/2} << 1$, hence the main part of reconnection is completed in the standard MHD regime. It is worth emphasizing that this transition from the Hall- to the standard MHD occurs when thickness of the CS is much smaller than the ion inertial length (moreover, the former reduces even further in the course of the subsequent standard MHD reconnection). Note also that these results confirm all basic conclusions reported in Ref.20, which, however, were relying on somewhat heuristic argumentation.

### III. Onset of plasmoid instability

According to Ref.[8], in the framework of the standard MHD the onset of plasmoid instability during forced magnetic reconnection is possible only when the amplitude of the external perturbation is large enough:

$$\delta/a > S^{-1/3}. \qquad (40)$$

In this case a central role is played here by the nonlinear equilibrium with the CS of thickness $\Delta \sim \delta$, (hereafter dimensional units are used), which is formed at the time $t = t_1 \sim \tau_A(\delta/a)^{-1} \ll t_S$ (in this Section it is assumed, for simplicity, that parameters $\varepsilon \equiv \alpha a$ and $ka$ are of the order of unity). Thus, consider first what a difference, if any, is made by the Hall-effect in the plasmoid instability of this CS. First of all, note that its very formation is due to the nonlinear torque in the vorticity Eq.(10)[], therefore, it can happen only before the Hall-effect comes into play, i.e. when $t_1 < t_H \sim \tau_A(d_i/a)^{-1}$ [see Eq.(24)], hence, it requires $d_i < \delta$. It turns out, however, that even under this restriction the Hall effect could be significant. In order to demonstrate it, one may find helpful a brief summary of the tearing instability theory in the standard MHD[16], and the Hall-MHD[14] frameworks, applied to a CS of thickness $l$, length $L \gg l$, and magnetic field $B$. These define respective Alfven velocity $V_A^{(l)}$, Alfven transit time $\tau_A^{(l)} = l/V_A^{(l)}$, and Lundquist number $S_l = lV_A^{(l)}/\eta$. Then, the standard MHD yields the instability growth rate

$$\gamma \tau_A^{(l)} \sim [S^{(l)}]^{-3/5}(ql)^{-2/5}, \qquad (41)$$

where $q$ is a wavenumber of the unstable tearing mode. The above expression is valid for a wave-length $\lambda = 2\pi/q$ in the interval $l < \lambda < \lambda_* \sim l S_l^{1/4}$ (it is assumed that $S_l \gg 1$). The growth rate falls sharply[21] when $\lambda > \lambda_*$, which makes such modes of no interest. Therefore, as seen from (41), the most unstable mode (the one with a maximum increment $\gamma$) corresponds to a wave-length $\lambda = \min\{\lambda_*, L\}$ (clearly, the CS of a finite length $L$ cannot accommodate perturbations with $\lambda > L$). Thus, it has the following implication to the nonlinear CS under consideration, for which $l \approx \delta, L \approx a, B \approx B_0(\delta/a)$, hence $V_A^{(l)} \approx V_A(\delta/a), \tau_A^{(l)} \approx \tau_A, S_l \approx S(\delta/a)^2$ [note that this $S_l \gg 1$ due to condition (40)]. Therefore, a ratio $\frac{\lambda_*}{a} \sim S^{1/4}\left(\frac{\delta}{a}\right)^{3/2}$, so the most unstable mode is that with $\lambda \sim \lambda_*$, if $(\delta/a) < S^{-1/6}$, or with $\lambda \sim a$ if otherwise. As pointed out in Ref.8, whatever the case, their growth rate is sufficient for the plasmoid instability development during the CS life-time $(\Delta t) \sim \delta^2/\eta \sim \tau_A S(\delta/a)^2$.

In the Hall-MHD the situation is even more favourable to the plasmoid instability development. Indeed, the Hall effect makes the secondary tearing instability faster by providing additional inflow of magnetic flux into the reconnection site, but leaves intact the CS resistive life-time $(\Delta t)$. Therefore, the question to answer now is how a finite value of $d_i$ affects the most unstable tearing mode, in particular, its wave-length. The latter is an import parameter which determines a number of plasmoids initially generated during the linear phase of the plasmoid instability. Thus, for the Hall-mediated tearing mode a summary, analogous to the one given above for the standard MHD case, reads as follows. For a mode with a wave-number $q$ transition to the Hall regime of instability occurs when

$$d_i/l > S_l^{-1/5}(ql)^{1/5}, \qquad (42)$$

which brings about the growth rate

$$\gamma \tau_A^{(l)} \sim S_l^{-1/2}(d_i/l)^{1/2}(ql)^{-1/2}. \qquad (43)$$

This expression holds for

$$l < \lambda < \lambda_*^{(H)} \sim l S_l^{1/3}(d_i/l)^{1/3}, \qquad (44)$$

and the growth rate falls sharply for $\lambda > \lambda_*^{(H)}$. In applying these results to the particular CS under consideration $[l \sim \delta, L \sim a, B \sim B_0(\delta/a)]$, one should also recall two constraints that are necessary for the very formation of this CS: $(\delta/a) > S^{-1/3}, d_i < \delta$. Thus, consider first the case when $S^{-1/3} < (\delta/a) < S^{-1/6}$, for which in the discussed above standard MHD framework a large number of the plasmoids is initially formed: $N_p \sim (a/\lambda_*) \sim S^{-1/4}(\delta/a)^{-3/2} > 1$. If a similar multi-plasmoid regime takes place in the Hall-MHD, the following two conditions must be met. Firstly, the optimal wave-length $\lambda_*^{(H)}$, given by Eq.(44), must be shorter than L, which in our case translates into

$$\lambda_*^{(H)} \sim \delta S^{1/3}\left(\frac{\delta}{a}\right)^{2/3}\left(\frac{d_i}{\delta}\right)^{1/3} < a \Rightarrow \frac{d_i}{\delta} < S^{-1}\left(\frac{\delta}{a}\right)^{-5}. \qquad (45)$$

Secondly, $d_i$ should be large enough to bring about the Hall-mediated reconnection [see Eq.(42)], hence

$$\frac{d_i}{\delta} > S^{-1/5}\left(\frac{\delta}{a}\right)^{-2/5}\left(\frac{\delta}{\lambda_*^{(H)}}\right)^{1/5} \Rightarrow \frac{d_i}{\delta} > S^{-1/4}\left(\frac{\delta}{a}\right)^{-1/2}.$$ These two inequalities are compatible if $(\delta/a) < S^{-1/6}$, while for $(\delta/a) < S^{-1/5}$ the validity of (45) is guaranteed due to the requirement of $d_i < \delta$. Therefore, in the Hall-MHD scenario the multi-plasmoid regime survives when $S^{-1/3} < (\delta/a) < S^{-1/5}$. Within the interval $S^{-1/5} < (\delta/a) < S^{-1/6}$ there are two possibilities. The multi-plasmoid case realizes if inequality (45) still holds, otherwise the most unstable mode is the one with $\lambda \sim a$, i.e. a number of the initially generated plasmoids is just a few. The latter is also the case when $(\delta/a) > S^{-1/6}$. Indeed, the Hall-reconnection condition (42) takes then the form

$$\frac{d_i}{\delta} > S^{-1/5}\left(\frac{\delta}{a}\right)^{-2/5}\left(\frac{\delta}{a}\right)^{1/5} \Rightarrow \frac{d_i}{\delta} > S^{-1/5}\left(\frac{\delta}{a}\right)^{-1/5},$$

(it guarantees that the inequality opposite to (45) holds), which is also compatible with the requirement $d_i < \delta$.

Thus, impact of the Hall effect on the plasmoid instability of the CS formed at the nonlinear stage of the ideal MHD evolution, is two-fold. The instability develops faster, and a wave-length of the most unstable mode becomes longer, the latter means a reduced number of the initially generated plasmoids. These changes, however, are not dramatic, as the overall scenario is basically the same as in the standard MHD case.

A very different situation is possible when the ion inertial length is large enough, so that the Hall-effect becomes instrumental during the entire resistive phase of the forced reconnection process. According to Section 2, this is the case when

$$d_i > aS^{-1/5}, \delta < d_i. \qquad (46)$$

The point is that in the standard MHD framework the plasmoid instability cannot develop at this stage: the system slips into the Rutherford regime of slow magnetic reconnection[8,22] even under quite a small perturbation amplitude. However, this effect is irrelevant in the Hall-MHD, where magnetic field is advected to the reconnection site by the Hall-generated electric current rather than by the plasma flow. Therefore, the situation becomes more favourable to the plasmoid instability development.

Thus, consider tearing stability of the CS formed at the major phase of the Hall-MHD forced reconnection [see Eqs.(34-35)] under conditions (46). In this case the CS parameters are as follows: $L \sim a, l \sim \Delta_H \sim aS^{-1/2}(d_i/a)^{-1/2}, B \sim B_0(\delta/a)$, which yield

$$V_A^{(l)} \sim V_A \frac{\delta}{a}, \tau_A^{(l)} \sim \frac{l}{V_A^{(l)}} \sim \tau_A S^{-1/2} \left(\frac{d_i}{a}\right)^{-1/2} \left(\frac{\delta}{a}\right)^{-1}, S_l = \frac{lV_A^{(l)}}{\eta} \sim S^{1/2} \left(\frac{d_i}{a}\right)^{-1/2} \left(\frac{\delta}{a}\right).$$

Then, according to (44), the wave-length of the most unstable tearing mode is equal to

$$\lambda_*^{(H)} \sim lS_l^{(1/3)} \left(\frac{d_i}{l}\right)^{1/3} \sim aS^{-1/6} \left(\frac{d_i}{a}\right)^{-1/6} \left(\frac{\delta}{a}\right)^{1/3}, \tag{47}$$

hence, $\lambda_*^{(H)} < a$ under the conditions (46). Therefore, this mode can develop in the CS of length $L \sim a$ and, hence, lead to a multiple-plasmoid $(N_p \sim a/\lambda_*^{(H)} > 1)$ initial phase of instability, provided that its growth rate is sufficiently high. In order to verify that the latter is the case, one should compare the respective increment $\gamma_*^{(H)}$ with the life-time $(\Delta t)$ of this CS, which in this case is the Hall reconnection time (33): $(\Delta t) \sim \tau_r^{(H)} \sim \tau_A S^{1/2}(d_i/a)^{-1/2}$. Thus, according to (43) and (47),

$$\gamma_*^{(H)} \sim [\tau_A^{(l)}]^{-1} S_l^{-1/3} \left(\frac{d_i}{l}\right)^{2/3} \sim \tau_A^{-1} S^{2/3} \left(\frac{d_i}{a}\right)^{5/3} \left(\frac{\delta}{a}\right)^{2/3}, \text{ which yields}$$

$$\gamma_*^{(H)} \cdot (\Delta t) \sim S^{7/6} \left(\frac{d_i}{a}\right)^{7/6} \left(\frac{\delta}{a}\right)^{2/3}. \text{ Therefore, the plasmoid instability requirement,}$$

$\gamma_*^{(H)} \cdot (\Delta t) > 1$, reads $\frac{\delta}{a} > S^{-7/4} \left(\frac{d_i}{a}\right)^{-7/4}$, that can be readily satisfied under the conditions (46).

## IV. Summary and discussion

The first part of the paper (Section II) presents detailed analytical theory of the Hall-MHD forced magnetic reconnection. The role of the Hall effect in this process is determined by the parameter $d \equiv d_i/a$. Thus, it is shown that there are two threshold values, $d_1 \sim S^{-1/3}$ and $d_2 \sim S^{-1/5}$, which separate different regimes of reconnection. If $d < d_1$, the Hall-effect plays no role at all, so the reconnection time follows the standard MHD scaling[7]: $\tau_r \sim \tau_A S^{3/5}$. In the intermediate case, when $d_1 < d < d_2$, initially reconnection proceeds in the Hall-MHD regime. However, it quickly gives way to the standard MHD phase, and the overall reconnection time still does not depend on the Hall parameter $d$. Only when the latter exceeds the second threshold, $d > d_2$, the Hall effect becomes dominant, and the reconnection time scales as $\tau_r = \tau_r^{(H)} \sim \tau_A S^{1/2} d^{-1/2}$.

Two relevant points are due here. The first one is about a widely accepted paradigm[23] that transition from the standard- to Hall-MHD occurs when the ion-inertial length $d_i$ exceeds the CS thickness $\Delta$. Our results clearly demonstrate that, generally speaking, this is not true. Thus, in the case of $d < d_1$, the standard MHD reconnection goes all the way even though $\Delta$ may become smaller than $d_i$ in the process. Moreover, consider the intermediate case, $d_1 < d < d_2$, when due to the CS shrinking during the ideal MHD phase a transition to the Hall-MHD does take place at the point when $\Delta \approx d_i$. Nevertheless, the system reverses back to the standard MHD evolution later on, despite the fact that at this stage $\Delta << d_i$.

The second point is concerning perturbation of the magnetic field component perpendicular to the reconnection plane, $B_z^{(1)}$. A part of it, $b_2$ [see Eq.(18)], has a quadrupole symmetry, and is commonly considered as a signature of the Hall-mediated magnetic reconnection[15]. However, it has been already pointed out[24] that the overall structure of $B_z^{(1)}$ could be more complicated. It is shown here that although this effect is weak in the case of a strong guide field, ($\varepsilon \equiv \alpha a << 1 \Rightarrow b_1 << b_2$), it could be significant when $\varepsilon \sim 1$, making then $b_1 \sim b_2$.

The second part (Section III) deals with the onset of plasmoid instability in the framework of Hall-MHD. As shown in Ref.8, in the standard MHD case the plasmoid instability becomes involved in the process of forced magnetic reconnection via the nonlinear CS forming at the ideal MHD stage of the system evolution. The main difference made here by the Hall-effect is a reduction in the number of initially generated plasmoids. This is because in the Hall-MHD the most unstable secondary tearing mode has a longer wavelength: according to Eq. (42), the Hall-effect has stronger impact on tearing perturbations with smaller wave-numbers $q$.

There is, however, another, more significant change: onset of plasmoid instability in the course of resistive evolution of the CS. This is not possible in the standard MHD, because the system slips into the Rutherford regime due to halting of the plasma flow. Favourably to the plasmoid instability development, this effect becomes irrelevant in the Hall-MHD regime, when advection of the poloidal magnetic field is provided by the Hall-generated electric current rather than by the bulk flow of the plasma. This enables a multi-plasmoid regime of the secondary tearing instability.

**Acknowledgments**

One of the authors, G.V., acknowledges financial support and warm hospitality at the Institute for Space-Earth Environmental Research, where the presented work has been completed. The work of K.K. was supported by MEXT/JSPS KAKENHI Grant JP15HO5814.